\documentclass{jfm}
\usepackage{graphicx}
\usepackage{epstopdf, epsfig}

\usepackage{amsmath,mathtools,amssymb,bm}
\usepackage{color}

\newcommand{\pfr}[2]{\ensuremath{\frac{\partial #1}{\partial #2}}}
\newcommand{\pfi}[2]{\ensuremath{{\partial #1}/{\partial #2}}}

\newcommand{\dd}{{\rm d}}

\newcommand{\beq}{\begin{equation}}
\newcommand{\eeq}{\end{equation}}

\newcommand{\Reyn}{\text{\textit{Re}}}
\renewcommand{\p}{\partial}

\catcode`\@=11
\def\gtsim{\mathrel{\vcenter{\m@th\offinterlineskip
\hbox{$\hfill>\hfill$}\kern.5ex\hbox{$\hfill\sim\hfill$}}}}
\catcode`\@=12

\catcode`\@=11
\def\ltsim{\mathrel{\vcenter{\m@th\offinterlineskip
\hbox{$\hfill<\hfill$}\kern.5ex\hbox{$\hfill\sim\hfill$}}}}
\catcode`\@=12

\newcommand{\tu}{\tilde u}
\newcommand{\tv}{\tilde v}
\newcommand{\tw}{\tilde w}
\newcommand{\tr}{\tilde r}
\newcommand{\tz}{\tilde z}

\newcommand{\tp}{\tilde \psi}

\newcommand{\hr}{\hat r}
\newcommand{\hz}{\hat z}
\newcommand{\hu}{\hat u}
\newcommand{\hw}{\hat w}

\shorttitle{A model for the constant-density boundary layer surrounding fire whirls}
\shortauthor{A.~D.~Weiss and others}

\title{A model for the constant-density boundary layer surrounding fire whirls}

\author{%
A.~D.~Weiss\aff{1},
P.~Rajamanickam\aff{1}\corresp{Current affiliation: Department of Aerospace
Engineering, Auburn University, Auburn, USA},
W.~Coenen\aff{1,2}\corresp{\email{wcoenen@ing.uc3m.es}},
A.~L.~S\'anchez\aff{1}\and
F.~A.~Williams\aff{1}}

\affiliation{%
\aff{1}Department of Mechanical and Aerospace Engineering, University of California San Diego, \\
La Jolla, USA
\aff{2}Grupo de Mec\'anica de Fluidos, Departamento de Ingenier\'ia T\'ermica y de Fluidos,
Universidad Carlos III de Madrid, Legan\'es (Madrid), Spain}

\begin{document}

\maketitle

\begin{abstract}
This paper investigates the steady axisymmetric structure of the cold boundary-layer flow surrounding fire whirls developing over localized fuel sources lying on a horizontal surface. The inviscid swirling motion found outside the boundary layer, driven by the entrainment of the buoyant turbulent plume of hot combustion products that develops above the fire, is described by an irrotational solution, obtained by combining Taylor's self-similar solution for the motion in the axial plane with the azimuthal motion induced by a line vortex of circulation $2 \pi \Gamma$. The development of the boundary layer from a prescribed radial location is determined by numerical integration for different swirl levels, measured by the value of the radial-to-azimuthal velocity ratio $\sigma$ at the initial radial location. As in the case $\sigma=0$, treated in the seminal boundary-layer analysis of  \citet{Burggraf.etal.1971}, the pressure gradient associated with the centripetal acceleration of the inviscid flow is seen to generate a pronounced radial inflow. Specific attention is given to the terminal shape of the boundary-layer velocity near the axis, which displays a three-layered structure that is described by matched asymptotic expansions. The resulting composite expansion, dependent on the level of ambient swirl through the parameter $\sigma$, is employed as boundary condition to describe the deflection of the boundary-layer flow near the axis to form a vertical swirl jet. Numerical solutions of the resulting non-slender collision region for different values of $\sigma$ are presented both for inviscid flow and for viscous flow with moderately large values of the controlling Reynolds number $\Gamma/\nu$. The velocity description provided is useful in mathematical formulations of localized fire-whirl flows, providing consistent boundary conditions accounting for the ambient swirl level.

\end{abstract}


\section{Introduction}
\label{sec:introduction}

\begin{figure}
\centering
\includegraphics[width=\textwidth]{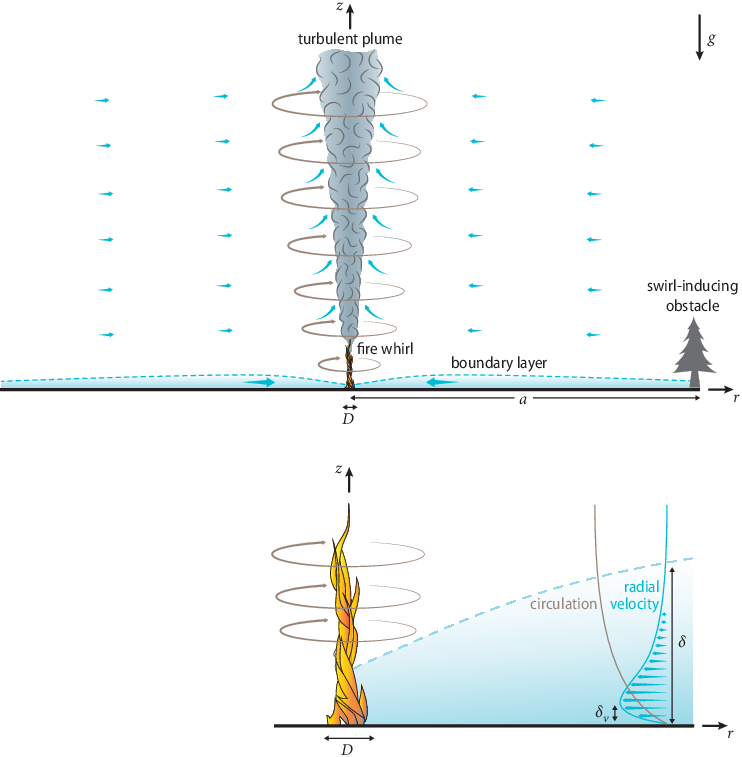}
\caption{General overview.}
\label{fig:overview}
\end{figure}

Fire whirls are vortical columns with a concentrated burning core. Observed
lengths vary from about 0.1 m in small experiments to tens of meters in wildland
fires. As stated in the recent review paper by \cite{Tohidi.etal.2018}, despite
significant research efforts, the current understanding of the flow structure
and dynamics of fire whirls, including the reasons for their dramatic
flame-lengthening effect and increased burning rate, is far from complete. The
present paper contributes to the needed understanding by investigating the
steady axisymmetric structure of the cold outer flow surrounding fire whirls
developing over localized fuel sources lying on a horizontal surface, a
configuration shown schematically in figure~\ref{fig:overview}. Attention will
be directed to the development of the important near-wall boundary layer and the
collision consequent to its radially inward flow component in the vicinity of
the fire whirl. The structures of the
plume and of the interior of the fire whirl depicted in the figure are not
analyzed here.

The flow of cold air surrounding the fire whirl, at distances much larger than the size of the fuel source (e.g. the diameter $D$ of the fuel pool in liquid-pool fires), ultimately is driven by the buoyant turbulent plume of hot combustion products that develops above the fire. As revealed by detailed experimental measurements \citep{lei2015temperature}, the temperature in fire-whirl plumes decays exponentially with radial distance from the axis towards the ambient value, so that density variations are only encountered near the axis, while the flow induced outside by the entrainment of the turbulent plume has constant density. Since the volumetric entrainment rate per unit length increases with the two thirds power of the vertical distance \citep{batchelor1954heat}, the effect of the slender plume on the outer flow is that of a semi-infinite line sink of varying strength, resulting in a self-similar potential solution described by \cite{Taylor.1958}. In the presence of obstacles, this meridional flow may be deflected, introducing an azimuthal velocity component, a fundamental ingredient in the development of fire whirls \citep{Tohidi.etal.2018} and other naturally occurring vortex phenomena, such as tornadoes \citep{Rotunno.2013} and dust devils \citep{maxworthy1973vorticity}. In wildland fires, for example, flow deflections beyond those associated with the circulation in weather patterns may be the result of flow interactions with topological features or tall vegetation, while in laboratory experiments on fire whirls and dust devils the deflection is achieved by surrounding the experimental setup with rotating circular screens \citep{Emmons.Ying.1967}, thin vertical flow vanes placed at a nonzero angle with respect to the radial direction \citep{Mullen.Maxworthy.1977,coenen2019observed}, or offset cylindrical or planar walls that leave small vertical slits for the tangential inflow of the incoming air \citep{Byram.Martin.1962}. 

The specific characteristics of the resulting inviscid swirling flow depend on the flow-deflection mechanism. For example, while the flow deflection by vertical vanes can be expected to be largely irrotational, the use of rotating circular screens may introduce a significant amount of azimuthal vorticity, which is not considered in the present analysis. In the present investigation, as in some laboratory experiments \citep{coenen2019observed}, the distance $a$ at which the circulation is induced is large compared with the flame height enhanced by whirl augmentation, so that the Taylor solution, applicable for turbulent plumes with sufficiently weak swirl, can be employed. Strong-swirl solutions that would generate different inviscid external flow fields at lower altitudes, which then would require a different analysis if $a$ were smaller, are not available. The production of swirl by deflection of the flow entrained by the turbulent plume is a distinctive characteristic of fire whirls, not present in swirl combustors, for instance, where the swirl is imparted prior to injection into the combustion chamber \citep{gupta1984swirl,candel2014dynamics}, leading to flow structures that are markedly different from those analyzed here.

The presence of swirl in the flow surrounding the fire whirl is accompanied by an increased radial pressure gradient, needed to balance the centripetal acceleration. Viscous forces decelerate the swirling motion in a near-wall boundary layer, where the imposed pressure gradient generates an overshoot of the radial inflow, which becomes more pronounced on approaching the axis. This flow feature was investigated in detail by \citet{Burggraf.etal.1971} for the specific case of a boundary layer on a fixed, non-rotating circular disk of radius $a$ whose axis is concentric with a potential vortex with circulation $2 \pi \Gamma$. Their analysis clarified in particular the structure of the terminal velocity profile found at small radial distances $r^* \ll a$, including a near-wall viscous sublayer of shrinking thickness $(\nu/\Gamma)^{1/2} r^*$ and a nearly inviscid layer of finite thickness $\delta=(\nu/\Gamma)^{1/2} a$, with $\nu$ representing the kinematic viscosity.

The boundary layer surrounding a fire whirl depends on the outer inviscid flow through its near-wall radial distributions of both azimuthal and radial velocity. To clarify the effect of the latter on the boundary-layer development, the previous potential-vortex analysis \citep{Burggraf.etal.1971}, in which the flow outside the boundary layer was purely azimuthal, is extended here by using as a model for the inviscid outer flow the potential solution obtained by combining linearly Taylor's potential solution \citep{Taylor.1958} for the flow in the axial plane with a potential vortex for the azimuthal motion. Numerical integrations of the boundary-layer equations are used to describe the development of the boundary layer for selected radial-to-azimuthal velocity ratios. A consistent asymptotic description is given for the  terminal velocity profiles at the axis, whose structure includes a thick external layer, additional to the two layers identified earlier by \citet{Burggraf.etal.1971}, which is needed to describe the transition to Taylor's radial flow. A composite expansion combining the results of the three layers in a single expression is developed for the profiles of radial and azimuthal velocity, providing an accurate description for the flow approaching the base of the fire whirl. 

As in the potential-vortex analysis \citep{Burggraf.etal.1971}, the radial mass flux carried by the wall boundary layer tends to a finite value on approaching the axis. The subsequent boundary-layer collision leads to the upward deflection of the flow in a nonslender region scaling with the characteristic near-axis boundary-layer thickness $\delta=(\nu/\Gamma)^{1/2} a$. Similar nonslender collision regions have been found in other buoyancy-driven flows, for instance in free convection from a heated sphere, where the eruption of the fluid into the plume above the sphere is the result of the collision of the boundary layer at the upper stagnation point, as described by \cite{potter1980free}. Because of its relevance in connection with tornados, its inviscid structure has been investigated in the past, using as lateral boundary condition the velocity profile induced by a potential vortex \citep{Fiedler.Rotunno.1986}. Additional results are presented below for fire whirls, with results given for different values of the ambient swirl, 
including profiles of vertical velocity for the deflected stream, which are ultimately responsible for the locally observed lengthening of fire-whirl flames. Furthermore, the validity of the inviscid description is critically assessed by investigating the accompanying boundary layer that develops near the wall in the collision region. Although boundary-layer separation is found to occur in all cases at a finite distance from the axis, additional integrations of the Navier-Stokes equations for moderately large values of the relevant Reynolds number $\Gamma/\nu$ reveal that the boundary layer reattaches before reaching the axis to form a slender recirculating bubble, so that the inviscid description remains largely valid.

\section{Boundary-layer development}
\label{BL}

\subsection{Preliminary considerations}

The cold flow surrounding fire whirls, to be described using cylindrical polar coordinates $(r^*,\theta,z^*)$ and associated velocity components $(u^*,v^*,w^*)$, is induced by the entrainment of the turbulent plume that extends vertically above the flame along the axis of symmetry. The volumetric entrainment rate per unit length, taken from investigations free from fire-whirl swirl, increases with the two thirds power of the vertical distance $z^*$ according to $\Phi= 2\pi C B^{1/3} z^{*2/3}$, where $B$ is the specific buoyancy flux~\citep{batchelor1954heat} and $C$ is a dimensionless factor, which approximately assumes the value $C=0.041$, as suggested by experimental results~\citep{Rouse.etal.1952,list1982turbulent}. Correspondingly, the flow induced in the axial plane has velocities decaying with the radial distance according to $C(B/r^*)^{1/3}$. For $r^* \gg [\nu/(CB^{1/3})]^{3/2}$ the associated Reynolds number $C B^{1/3} r^{*2/3}/\nu$ is large, resulting in nearly inviscid motion, which, in the absence of swirl, is described by a
self-similar potential solution that is due to  \cite{Taylor.1958}. The corresponding slip velocity at the wall is given by
\beq 
u^*_w=-A_T C (B/r^*)^{1/3}, \label{uw}
\eeq
involving the numerical factor 
\beq
A_T= \frac{4 (2)^{1/3} \pi^2}{3 \, \mathup{\Gamma}^3(1/3)} \simeq 0.8624,
\eeq
where $\mathup{\Gamma}$ denotes the Gamma function. The potential solution fails in the boundary layer, where the radial velocity, also self-similar, is given by $u^*/u^*_w=f_{\scriptscriptstyle{\rm T}}'$ in terms of the derivative of the reduced stream function $f_{\scriptscriptstyle{\rm T}}(\varsigma)$, a function of the rescaled vertical distance $\varsigma=(A_T C B^{1/3}/\nu)^{1/2} r^{* -2/3} z^*$ determined from the boundary-value problem 
\beq \label{fT_eq}
f_{\scriptscriptstyle{\rm T}}'''-\frac{4}{3}f_{\scriptscriptstyle{\rm T}} f_{\scriptscriptstyle{\rm T}}''+\frac{1}{3} (1-f_{\scriptscriptstyle{\rm T}}^{\prime 2})=0; \quad f_{\scriptscriptstyle{\rm T}}(0)=f_{\scriptscriptstyle{\rm T}}'(0)=f_{\scriptscriptstyle{\rm T}}'(\infty)-1=0,
\eeq
the subscript T referring to the boundary layer accompanying Taylor's potential flow. In the notation employed throughout the paper the prime denotes differentiation of functions of one variable (e.g. in the above description, it represents differentiation with respect to the self-similar coordinate $\varsigma$). It is worth mentioning that the flow structure surrounding turbulent plumes, relevant to fire whirls, is fundamentally different from that surrounding laminar plumes, characterized by small entrainment rates $\Phi \sim \nu$ and associated flow Reynolds numbers of order unity, the case analyzed by \cite{Schneider.1981}, who found an exact self-similar solution of the first kind for the swirl-free flow in the axial plane. As shown recently by \citet{Coenen.etal.2019}, the accompanying circulation in this viscous case is described by a self-similar solution of the second kind.

As previously mentioned, the three-dimensional inviscid motion surrounding fire whirls is affected by the manner in which swirl is imparted to the flow. Instead of focusing on a specific configuration, for generality in the following analysis the inviscid flow outside the boundary layer will be described using as a canonical model the exact axisymmetric solution of the Euler equations resulting from combining Taylor's potential solution for the meridional flow and a line vortex of circulation $2 \pi \Gamma$ for the azimuthal flow. Correspondingly, at the outer edge of the boundary layer the radial velocity approaches the value given in~\eqref{uw} while the azimuthal component approaches the value
\beq 
v^*_w=\Gamma/r^*. \label{vw}
\eeq
The presence of swirl alters the flow across the boundary layer, so that, even for this model
problem, a self-similar description, which is available in the absence of swirl, as described
above in~\eqref{fT_eq}, does not exist when $v^*_w \ne 0$. The fundamental lack of flow
similarity can be illustrated by considering the flow at large radial distances, where the
radial motion becomes dominant, as can be inferred from the different decay rates present
in~\eqref{uw} and~\eqref{vw}. Correspondingly, as $r^* \rightarrow \infty$ the self-similar
function $u^*/u^*_w=f_{\scriptscriptstyle{\rm T}}'(\varsigma)$ appears to be the appropriate
leading-order representation for the radial velocity across the boundary layer, while the
azimuthal velocity $v^*=\Gamma g_{\scriptscriptstyle{\rm T}}(\varsigma)/r^*$ should be
determined by the accompanying problem
\beq 
g_{\scriptscriptstyle{\rm T}}''-\frac{4}{3} f_{\scriptscriptstyle{\rm T}} g_{\scriptscriptstyle{\rm T}}'=0; \quad g_{\scriptscriptstyle{\rm T}}(0)=g_{\scriptscriptstyle{\rm T}}(\infty)-1=0,
\eeq
obtained at leading order from the axisymmetric boundary-layer form of the azimuthal momentum equation. This presumed self-similar structure fails, however, because the last problem has no solution, which can be seen by investigating the behavior as $\varsigma \to \infty$ of the first integral $g_{\scriptscriptstyle{\rm T}}'=g_{\scriptscriptstyle{\rm T}}'(0) \exp\left[\frac{4}{3} \int_0^\varsigma  f_{\scriptscriptstyle{\rm T}} {\rm d} \tilde{\varsigma} \right]$ to show that $g_{\scriptscriptstyle{\rm T}}'(0)=0$ to avoid divergence, so that the only possible solution is $g_{\scriptscriptstyle{\rm T}}=$ constant, which cannot satisfy simultaneously both boundary conditions $g_{\scriptscriptstyle{\rm T}}(0)=g_{\scriptscriptstyle{\rm T}}(\infty)-1=0$; this lack of similarity is also encountered when the outer flow is driven solely by a potential vortex~\citep{gol1960paradoxical} (see also~\cite{king1964boundary} for a discussion of boundary-layer selfsimilarity when the outer azimuthal velocity varies with a general power of the radial distance).

Progress in understanding can be achieved by investigating the development of the boundary layer from a given radial location, as was done in the previous analysis of the boundary layer on a disk of radius $a$~\citep{Burggraf.etal.1971}. The same approach is to be considered below, with the ratio of the radial-to-azimuthal velocity 
\beq
\sigma=\frac{u^*_w(a)}{v^*_w(a)}
\eeq 
at the disk edge arising as the only controlling parameter in the resulting description. This
idealized disk problem may, for example, be considered to provide an approximate description of the main features of the boundary-layer flow in the region between the fire and swirl-producing vanes at radius $a$ in laboratory fire-whirl experiments, with the velocity ratio $\sigma$ being directly related to the angle of inclination of the vanes. In particular, the terminal velocity profile at $r^* \ll a$ can be anticipated to provide a realistic representation for the flow surrounding localized fire whirls, with the parameter $\sigma$ measuring the level of swirl introduced by the collective effect of the flow-deflecting obstacles, located at radial distances much larger than the characteristic size of the fuel source feeding the fire. 

\subsection{Problem formulation}

Following~\citet{Burggraf.etal.1971}, the problem is scaled using $a$ and 
\beq \label{delta_def}
\delta=\frac{a}{\sqrt{\Reyn}}
\eeq 
for the radial and axial coordinates, with
\beq \label{Re_def}
\Reyn=\frac{\Gamma}{\nu}
\eeq
representing the relevant Reynolds number. Correspondingly, the azimuthal and radial velocity components $u^*$ and $v^*$ are scaled with $\Gamma/a$, corresponding to a radial pressure gradient scaling with $\rho \Gamma^2/a^3$ ($\rho$ representing the density), while the axial component $w^*$ is scaled with $(\Gamma/a)/\sqrt{\Reyn}$, resulting in the dimensionless variables 
\begin{equation}
    r = \frac{r^*}{a}, \quad z= \frac{z^*}{\delta}, \quad u = \frac{u^*}{\Gamma/a}, \quad v = \frac{v^*}{\Gamma/a}, \quad w= \frac{w^*}{\Gamma/(a \sqrt{\Reyn})}.
\end{equation}
Neglecting terms of order $\Reyn^{-2} \ll 1$ reduces the conservation equations, written in their steady axisymmetric form for a constant-density fluid, to their boundary-layer form
\begin{align}
    u\pfr{u}{r} + w\pfr{u}{z} - \frac{v^2}{r} = -\frac{\sigma^2}{3r^{5/3}} -\frac{1}{r^3}  + \pfr{^2 u}{z^2}, \label{BL1}\\
    u\pfr{}{r} (rv) + w\pfr{}{z} (rv) = \pfr{^2}{z^2}(rv), \label{BL2}\\
    \pfr{}{r}(ru) + \pfr{}{z}(rw) =0. \label{BL3}
\end{align}
The first two terms on the right-hand side of ~\eqref{BL1} arise from the radial pressure gradients imposed by the external Taylor and potential-vortex flows, respectively. These equations are to be integrated for decreasing values of $r$ with the boundary conditions
\beq \label{farfield}
    u \rightarrow -\sigma r^{-1/3}, \quad v\rightarrow \frac{1}{r} \quad \text{as} \quad z\to \infty
\eeq
and
\beq
    u=v=w=0 \quad \text{at} \quad z=0 \label{BL_bc2}
\eeq
for $r < 1$ and the initial velocity profiles $u=-\sigma$ and $v=1$ at $r=1$, consistent with~\eqref{farfield}. The only parameter in the description is the initial flow inclination $\sigma$. As expected, for $\sigma=0$ the problem reduces exactly to that addressed by~\cite{Burggraf.etal.1971}.

\subsection{Sample numerical results}

\begin{figure}
\centering
\includegraphics[width=0.8\textwidth]{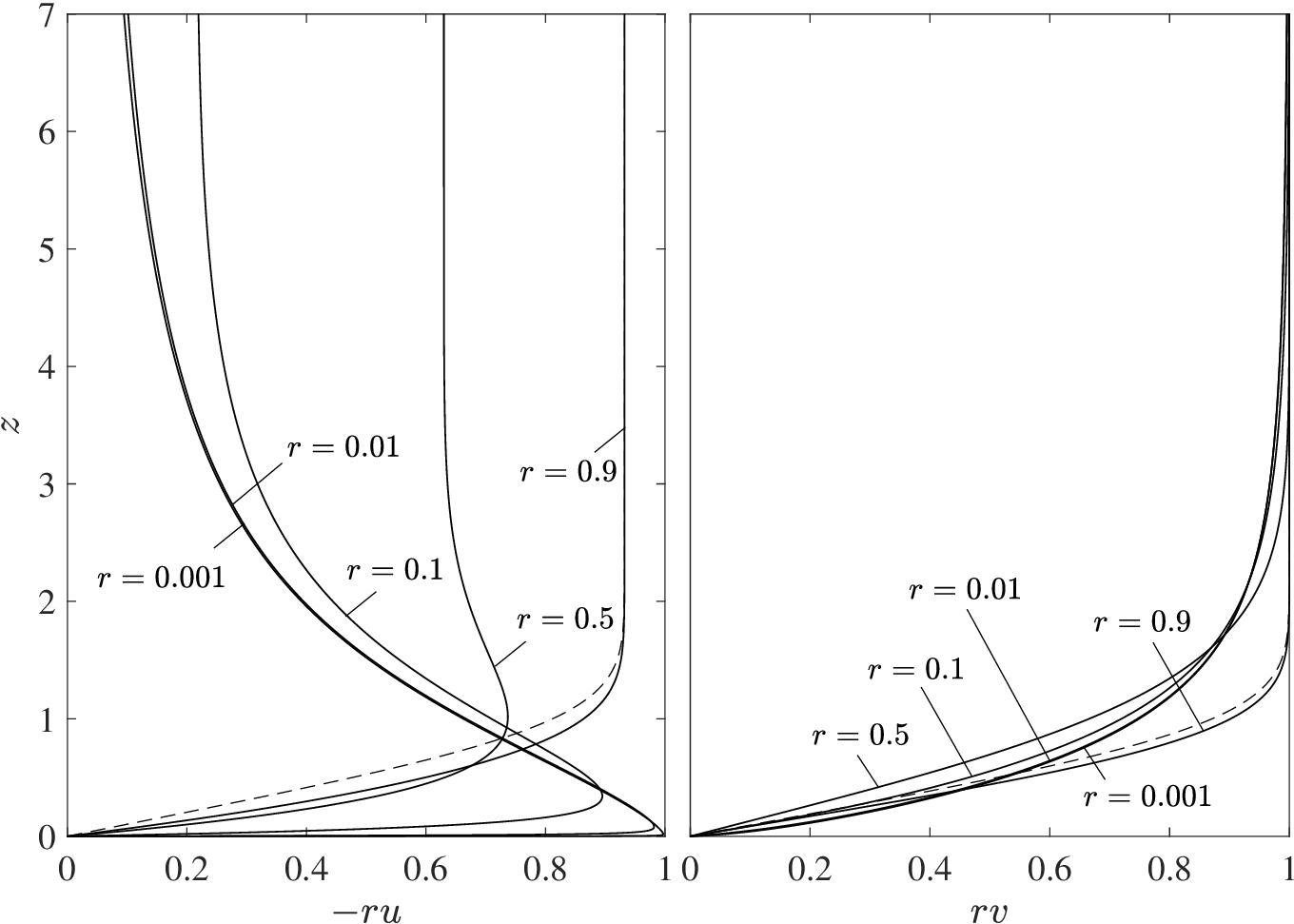}
\caption{Boundary-layer profiles of radial and azimuthal velocity at various radial locations for $\sigma=1$. The dashed curves represent the asymptotic predictions $-u/(\sigma r^{-1/3})=r v=f_{\scriptscriptstyle{\rm B}}'(\zeta)$ for $1-r \ll 1$ evaluated at $r=0.9$.}
\label{fig:boundarylayer}
\end{figure}

The problem~\eqref{BL1}--\eqref{BL_bc2} was integrated numerically by marching from $r=1$ with decreasing values of $r$ for selected values of $\sigma$. Sample results are shown in figure~\ref{fig:boundarylayer} for $\sigma=1$. The radial and azimuthal velocities are uncoupled for $1-r \ll 1$, when the effects of the centripetal acceleration $-v^2/r$ and pressure gradient $-\sigma^2 r^{-3/5}/3-r^{-3}$ can be neglected in~\eqref{BL1} at leading order, reducing the solution with $\sigma \ne 0$ to $-u/(\sigma r^{-1/3})=r v=f_{\scriptscriptstyle{\rm B}}'(\zeta)$, where $f_{\scriptscriptstyle{\rm B}}$ is Blasius stream function, obtained by integration of $f_{\scriptscriptstyle{\rm B}}'''+f_{\scriptscriptstyle{\rm B}} f_{\scriptscriptstyle{\rm B}}''/2=0$ with boundary conditions $f_{\scriptscriptstyle{\rm B}}(0)=f_{\scriptscriptstyle{\rm B}}'(0)=f_{\scriptscriptstyle{\rm B}}'(\infty)-1=0$, with the prime denoting here differentiation with respect to the local self-similar coordinate $\zeta=z/(1-r)^{1/2}$. The asymptotic predictions for $1-r \ll 1$ are compared in figure~\ref{fig:boundarylayer} with the profiles obtained numerically at $r=0.9$.

\begin{figure}
\centering
\includegraphics[width=0.8\textwidth]{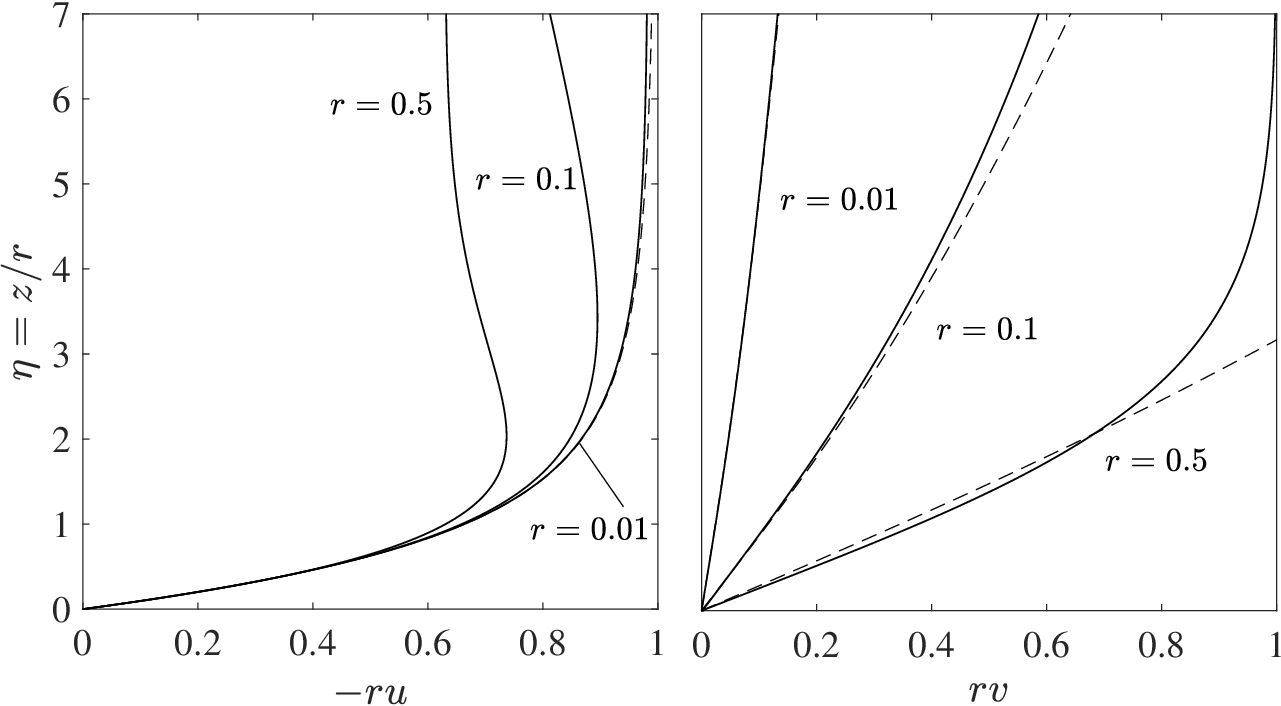}
\caption{Comparison of the near-wall boundary-layer profiles obtained from the asymptotic predictions $ru = - \psi_0'(\eta)$ and $rv=C_1 r^\lambda \gamma_1(\eta)$ (dashed curves) with those determined numerically at various radial locations by integration of~\eqref{BL1}--\eqref{BL3} for $\sigma = 1$ (solid curves).}
\label{fig:boundarylayer2}
\end{figure}

The effect of the azimuthal motion on the radial flow is no longer negligible as $1-r$ increases to values of order unity, leading to an overshoot in the radial velocity, as is already evident in the results of figure~\ref{fig:boundarylayer} for $r=0.5$. This overshoot becomes more prominent as $r \rightarrow 0$, with the peak value of $r u$ reaching a near-unity value at small distances $z \sim r$. A detailed view of this near-wall region is shown in figure~\ref{fig:boundarylayer2}, where the dashed curves represent analytic results, to be developed below. 

The profiles of $r u$ and $r v$ shown in figure~\ref{fig:boundarylayer} are seen to approach a terminal shape as $r \rightarrow 0$. Although the specific shape of these terminal profiles depends on the value of $\sigma$, all solutions show a common multi-layered asymptotic structure, which, with the exception of the outermost layer, is fundamentally similar to that described in \cite{Burggraf.etal.1971} for the potential vortex ($\sigma=0$). A detailed analysis of the different layers is given below, and the associated asymptotic solutions are combined to generate composite expansions for the radial and azimuthal velocity components, providing an accurate boundary-layer description for $r \ll 1$ (i.e. dimensional distances $r^* \ll a$).

\section{The structure of the terminal velocity profile}
\label{sec:terminal}

We consider now the solution to~\eqref{BL1}--\eqref{BL3} with boundary conditions \eqref{farfield} and \eqref{BL_bc2} in the asymptotic limit $r \ll 1$ for $\sigma \sim 1$. As noted by~\cite{Burggraf.etal.1971}, at leading order viscous effects are confined to a thin layer $z \sim r$, outside of which the flow is inviscid, with values of $-ru \sim 1$ and $1-rv \sim 1$ at distances $z \sim 1$. Unlike the potential-vortex solution $\sigma=0$, which exhibits velocity profiles with a rapid exponential decay away from the wall, in fire whirls the transition to the outer solution $ru=- \sigma r^{2/3}$ and $rv=1$ occurs in a fairly large external layer, which necessitates a separate analysis, as shown below, exercising the full formalism of matched asymptotic expansions \citep{lagerstrom2013matched}.

\subsection{The lower viscous sub-layer}

At leading order, the solution in the viscous sub-layer, where $rv \ll -ru$ (as is apparent from figure~\ref{fig:boundarylayer2}), is independent of $\sigma$ and corresponds to that described by~\cite{Burggraf.etal.1971}. With circulation neglected, the boundary--layer equation \eqref{BL1} can be expressed in terms of the self-similar coordinate $\eta = {z}/{r}$ and accompanying stream function $\psi = r \psi_0(\eta)$, defined such that $ru = - \psi_0'$ and $rw=\psi_0-\eta \psi'_0$, to give at leading order the problem
\begin{equation} \label{psi_0eq}
\psi_0''' - \psi_0 \psi_0'' - \psi_0'^2 + 1 =0; \qquad \psi_0(0)=\psi_0'(0)=\psi_0'(\infty) -1 =0.
\end{equation}
The solution, expressible in terms of the parabolic cylinder functions~\citep{Mills.1935}, provides the asymptotic behaviour $\psi_0 \simeq \eta-1.0864$ and
\begin{equation}
rw \rightarrow -1.0864 \label{rw_inner}
\end{equation}
for $\eta \gg 1$. The accompanying weak azimuthal motion is described by a self-similar solution of the second kind of the form $rv \propto r^\lambda \gamma_1(\eta)$, where the eigenfunction $\gamma_1(\eta)$ obeys the linear equation
\beq
  \gamma_1'' - \psi_0\gamma_1' + \lambda \psi_0'\gamma_1 = 0 
\eeq
stemming from~\eqref{BL2}. A nontrivial solution satisfying the non-slip condition $\gamma_1=0$ at $\eta=0$ and exhibiting algebraic growth $\gamma_1 \propto \eta ^\lambda$ (as opposed to exponential growth) as $\eta \rightarrow \infty$, as needed to enable matching with the outer profile, exists only for a discrete set of values of the eigenvalue $\lambda$, with the smallest eigenvalue, corresponding to the dominant eigenfunction for small $r$, found to be $\lambda = 0.6797$. 

The solution to the above eigenvalue problem determines the near-wall azimuthal velocity $r v = C_1 r^\lambda \gamma_1(\eta)$, aside from a multiplicative factor $C_1$, a function of $\sigma$ to be determined by matching with the outer inviscid solution. For definiteness, it is convenient to use $\gamma_1 = \eta ^\lambda$ as $\eta \gg 1$ as the normalization condition for the eigenfunction $\gamma_1$, resulting in the asymptotic behaviour
\beq \label{rvC1}
r v \rightarrow C_1 z^\lambda \quad {\rm as} \quad \eta \rightarrow \infty,
\eeq
to be employed below in the matching procedure. The near-wall asymptotic predictions $ru = - \psi_0'(\eta)$ and $r v = C_1 r^\lambda \gamma_1(\eta)$ for $r \ll 1$ are compared in figure~\ref{fig:boundarylayer2} with the results of numerical integration for $\sigma=1$. The predictions for $r v = C_1 r^\lambda \gamma_1(\eta)$ are computed with $C_1=0.5496$, the value obtained below by matching with the outer inviscid results when $\sigma=1$. The asymptotic predictions and the numerical results are seen to be virtually indistinguishable for $r=0.01$ in figure~\ref{fig:boundarylayer2}. 

\subsection{The main inviscid layer} 

\begin{figure}
\centering
\includegraphics[width=0.8\textwidth]{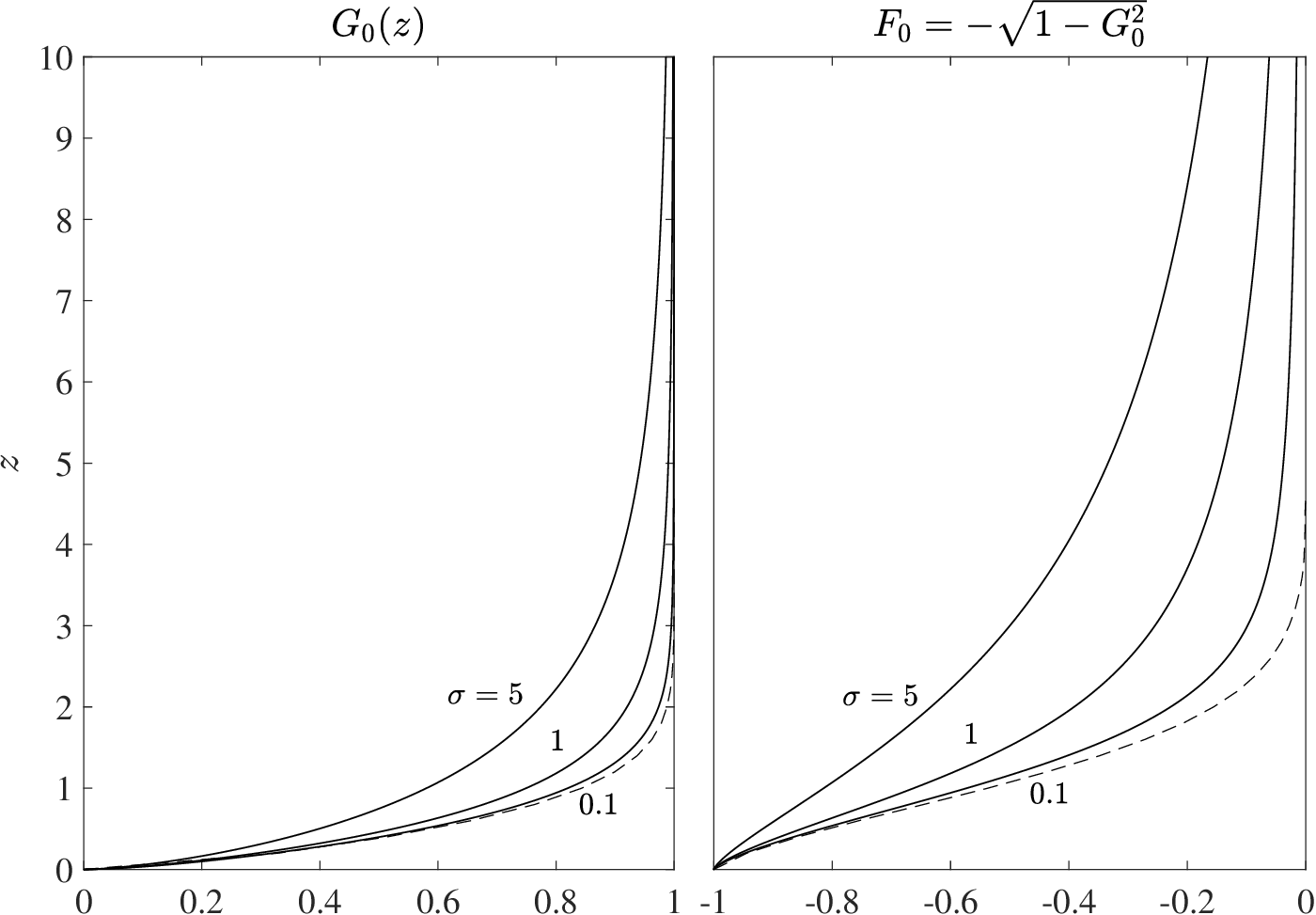}
\caption{Profiles of $G_0$ and $F_0$ for various $\sigma$. Shown as dashed line are the profiles of \cite{Burggraf.etal.1971} for $\sigma =0$.}
\label{fig:G0F0}
\end{figure}

In the intermediate layer $z \sim 1$, the expansions for the velocity components at $r \ll 1$ take the form
\beq \label{outerexpansion}
    ru= F_0(z) + rF_1(z) +\cdots, \quad
    rv= G_0(z)  + rG_1(z) + \cdots,\quad
    rw=  H_1(z) + \cdots. 
\eeq
The leading-order functions $F_0(z)$ and $G_0(z)$ must satisfy $F_0 \rightarrow -1$ and $z^{-\lambda} G_0 \rightarrow C_1$ as $z \rightarrow 0$, corresponding to matching with the viscous sublayer, and as $z \rightarrow \infty$ they must approach the outer values $F_0=G_0-1=0$, consistent at this order with the velocity found outside the boundary layer. The two functions $F_0$ and $G_0$ are related by the equation
\begin{equation} \label{eulersol}
    F_0^2 + G_0^2=1,
\end{equation}
which follows from the leading-order $1/r^3$ terms in~\eqref{BL1}, but, other than that, their specific shape depends on the development of the boundary layer for $0< r < 1$, yielding different profiles for different values of $\sigma$. The additional functions $F_1$, $G_1$, and $H_1$ appearing in~\eqref{outerexpansion} are related to the leading-order functions by
\beq
H_1=1.0864 F_0, \: F_1=- 1.0864 F'_0, \; G_1=- 1.0864 G'_0
\eeq
as can be seen by carrying the asymptotic solution to a higher order, with the numerical factor $1.0864$ selected to ensure inner matching with the vertical velocity~\eqref{rw_inner}.  

To determine $G_0$ and $F_0=-\sqrt{1-G_0^2}$, the numerical integration of~\eqref{BL1}--\eqref{BL3} was extended to extremely small radial distances $r \sim 10^{-4}$, and the asymptotic predictions $ru= F_0(z)- 1.0864 r F'_0(z)$ and $rv= G_0(z)- 1.0864 r G'_0(z)$ were used to extrapolate the result to $r=0$. The solution was further corrected to remove the viscous sublayer by replacing the solution at $z \ll 1$ with the near-wall behavior
 \beq \label{smallz}
     F_0 = -1 + \frac{1}{2} C_{1}^2 z^{2\lambda}, \quad G_0 = C_{1} z^{\lambda},
\eeq
arising from matching with~\eqref{rvC1}, with the constant $C_1$ obtained from the numerical integrations by evaluating $z^{-\lambda} r v$ at small distances $z \sim r$ from the wall, yielding for instance $C_1=(0.7125,0.8275,0.9065,0.9596,1.0181,1.6187)$ for $\sigma=(5,2,1,0.5,0.2,0.1)$. The observed evolution for decreasing values of $\sigma$ appears to be in agreement with the limiting value $C_1=1.6518$ reported by~\cite{Burggraf.etal.1971} for $\sigma=0$.


While the asymptotic behaviour $G_0=\sqrt{1-F_0^2} \propto z^{\lambda}$ as $z \rightarrow 0$ shown in~\eqref{smallz}  applies for both $\sigma=0$ and $\sigma \ne 0$, the solution as $z \rightarrow \infty$ is qualitatively different in these two cases, with the exponential decay found by~\cite{Burggraf.etal.1971} for $\sigma =0$ being replaced for $\sigma \ne 0$ by an algebraic decay of the form
\beq \label{largez}
    F_0 = -C_{2} z^{-\mu} + \cdots, \quad G_0 = 1 - \frac{1}{2} C_{2}^2 z^{-2\mu} + \cdots,
\eeq
where the factor $C_2$ and the exponent $\mu$ depend on $\sigma$. Their values were obtained from the numerical results by examining the decay with vertical distance of the near-axis terminal profiles of $ru$ and $rv$, yielding for instance $C_2=(2.85,1.50,0.88,0.47)$ and $\mu=(1.21,1.19,1.16,1.11)$ for $\sigma=(5,2,1,0.5)$. At a given $r \ll 1$, the range of $z$ over which~\eqref{largez} applies decreases for decreasing $\sigma$, thereby hindering the precise evaluation of $C_2$ and $\mu$ for $\sigma < 0.5$. The observed evolution of the approximate values, computed with use of the velocity profiles at the smallest radial distance reached in the boundary-layer computations (i.e. $r \simeq 10^{-4}$), indicates that the exponent $\mu$ decreases with decreasing $\sigma$ to approach unity as $\sigma \rightarrow 0$, with the accompanying value of $C_2$ vanishing in this limit.

\subsection{The upper transition region}

The asymptotic expansion~\eqref{outerexpansion} fails in a transition region corresponding to $z\sim r^{-2/(3\mu)} \gg 1$ where, according to~\eqref{largez}, the value of $F_0$ becomes of order $F_0 \sim r^{2/3}$, comparable to the limiting value $ru=-\sigma r^{2/3}$ found at $z=\infty$. This transition region can be described in terms of the order-unity similarity coordinate $\xi=(\sigma/C_2)^{1/\mu}r^{\frac{2}{3\mu}}z$ and associated rescaled velocity variables
\begin{align}
    ru= \sigma r^{2/3} f(\xi), \; rv= 1 + \sigma^2 r^{4/3}g(\xi), \; rw= C_2^{1/\mu}\sigma^{\frac{\mu-1}{\mu}}r^{-\frac{2+\mu}{3\mu} } h(\xi).
\end{align}
At leading order, the boundary-layer equations~\eqref{BL1}--\eqref{BL3} in this inviscid outer transition region simplify to
\begin{align}
\left(\frac{2\xi}{3\mu}f + h\right)f' &= \frac{f^2-1}{3} + 2g, \label{feq}\\
\left(\frac{2\xi}{3\mu}f + h\right)g' &= -\frac{4fg}{3}, \label{geq}\\
\left(f + \frac{\xi}{\mu}f'\right) + \frac{3h'}{2} &= 0. \label{weq}
\end{align}
The solution can be reduced to a quadrature as follows. Dividing \eqref{feq} by \eqref{geq} and integrating the resulting equation using the boundary conditions $f(\infty)\rightarrow -1$ and $g(\infty)\rightarrow 0$, which follow from matching with the outer potential solution, provides 
\beq \label{goff}
g = \frac{1}{2} (1-f^2). 
\eeq
Substitution of this result into~\eqref{feq} and elimination of $h$ with use of~\eqref{weq} leads to the autonomous equation
\begin{equation}
    (f^2 - 1)f'' - \frac{\mu+1}{\mu} ff'^2 =0,
\end{equation}
which can be integrated once using the boundary condition $f \rightarrow -\xi^{-\mu}$ as $\xi \rightarrow 0$, obtained by matching with~\eqref{largez}, to give $f'=\mu (f^2-1)^{(\mu+1)/(2\mu)}$, finally yielding
\begin{equation}
   \mu \xi = \int_{-\infty}^f \frac{\dd \tilde{f}}{(\tilde{f}^2-1)^{\frac{\mu+1}{2\mu}}},  \label{fsol}
\end{equation}
with $\tilde{f}$ being a dummy integration variable. The above integral, which is expressible in terms of incomplete beta function, provides, together with the previous equation~\eqref{goff}, the radial and azimuthal velocity distributions $f(\xi)$ and $g(\xi)$. 

Inspection of~\eqref{fsol} reveals that, for the values $\mu>1$ that apply in our description, the function $f$ is a front solution that reaches the boundary value $f=-1$ at a finite location $\xi=\xi_o$ given by
\begin{equation} \label{xi_o}
    \xi_o = \frac{1}{2\mu}{\rm B}\left(\frac{1}{2\mu}, \frac{\mu-1}{2\mu}\right), 
\end{equation}
as follows directly from~\eqref{fsol}, with ${\rm B}$ representing here the beta function. Note that, since the value of $\mu-1$ remains relatively small for $\sigma \sim 1$, the front is always located at large distances $\xi_o \simeq 1/(\mu-1)$.

\subsection{The composite expansion}

\begin{figure}
\centering
\includegraphics[width=0.80\textwidth]{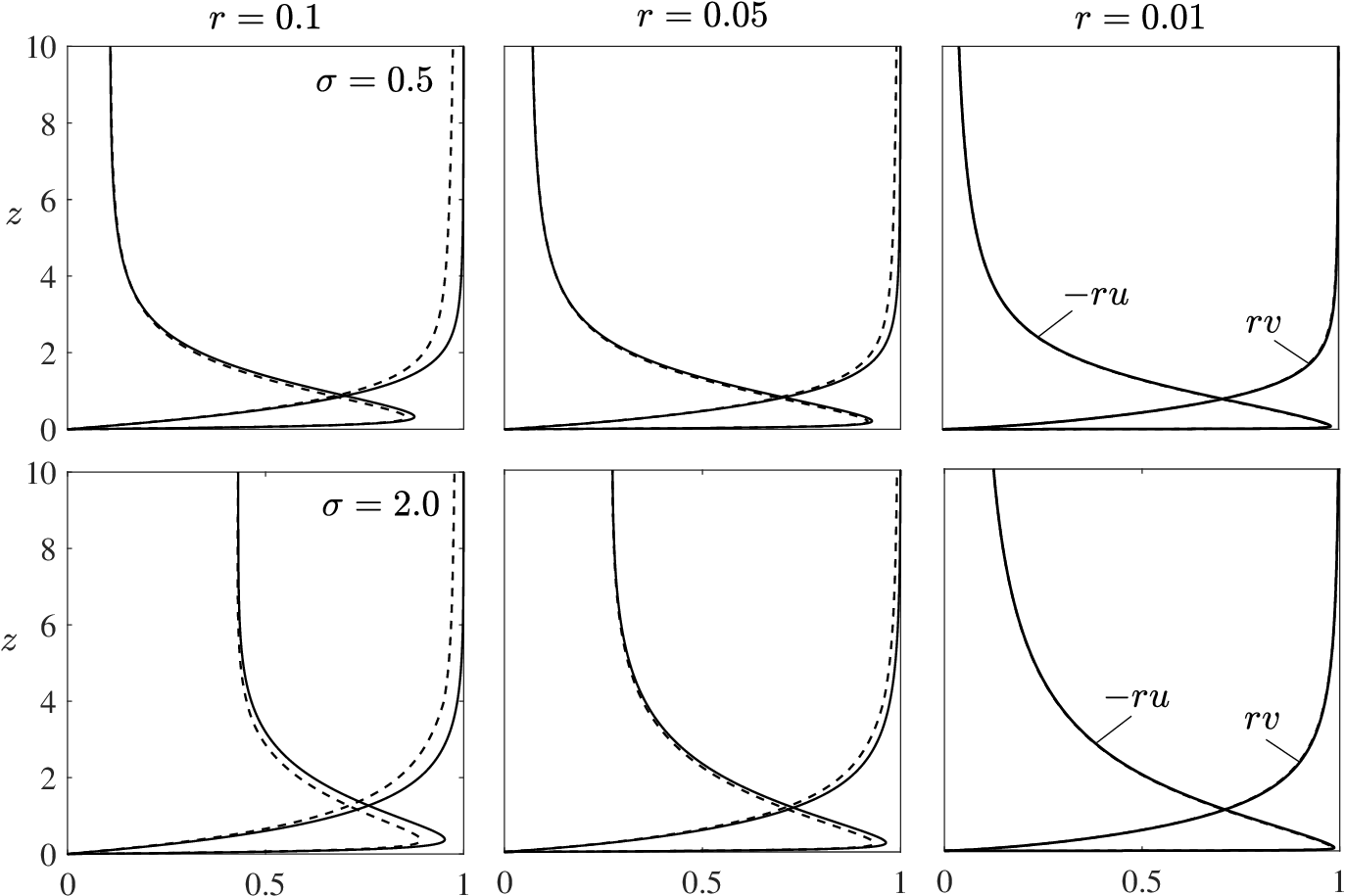}
\caption{Profiles of radial and azimuthal velocity for $\sigma =0.5$ (top row) and $\sigma = 2.0$ (bottom row) obtained at different radial locations $r\ll1$ by numerical integration of~\eqref{BL1}--\eqref{BL3} (solid curves) and by evaluation of the composite expansion \eqref{thecomposite} (dashed curves).}
\label{fig:composite}
\end{figure}

\label{subsub:composite}

The separate solutions found in the different regions can be used to generate the composite expansions 
\beq \label{thecomposite}
\begin{split}
ru &= - \psi^{\prime}_0(z/r) + 1 + F_0(z) + \sigma r^{2/3} f[(\sigma/C_2)^{1/\mu}r^{\frac{2}{3\mu}}z] + C_2 z^{-\mu},\\
rv &= C_1 r^\lambda \gamma_1(z/r) - C_1 z^\lambda + G_0(z) + \sigma^2 r^{4/3} g[(\sigma/C_2)^{1/\mu}r^{\frac{2}{3\mu}}z] + \tfrac{1}{2} C_2^2 z^{-2\mu},
\end{split}
\eeq
which describe the radial and azimuthal velocity profiles as $r\to 0$ with small errors of order $r$. The accuracy of these expansions is tested in figure~\ref{fig:composite} by comparing the asymptotic predictions with the results of numerical integrations of the boundary-layer equations~\eqref{BL1}--\eqref{BL3}. The degree of agreement displayed in the figure is clearly satisfactory, with the composite expansion being virtually indistinguishable from the numerical results at $r=0.01$.

\section{The collision region}
\label{sec:collision}

\begin{figure}
\centering
\includegraphics[width=\textwidth]{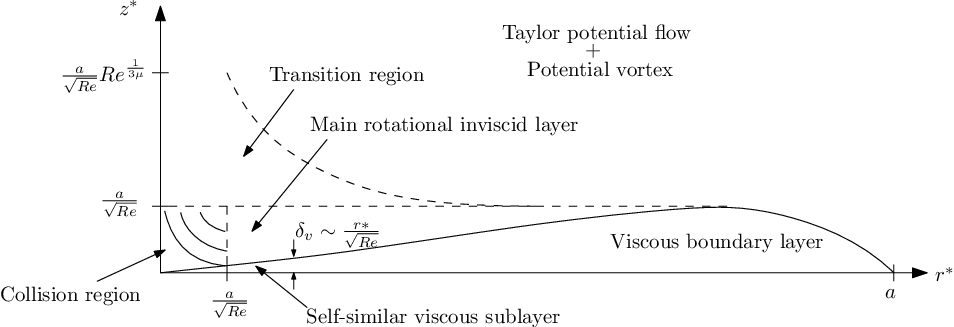} 
\caption{A schematic view of the boundary-layer flow with indication of the different scales.}
\label{fig:schematic_blflow}
\end{figure}

The boundary-layer flow, approaching the axis with a velocity nearly parallel to the wall, undergoes a rapid upward deflection in a non-slender collision region of characteristic size $\delta=a/\sqrt{\Reyn}$. This region is illustrated in figure~\ref{fig:schematic_blflow}, where the three-level flow outside that region also is shown. The thickness of the viscous sublayer decreases linearly with decreasing radius, the rotational inviscid layer occupying an increasing fraction of the initially non-similar boundary layer that emerges at the outer edge of the disk. For the incoming flow, described by the previous composite expansion, the viscous sublayer, the main inviscid layer, and the region of transition to potential flow at radial distances of order $r^* \sim \delta$ have associated thicknesses of increasing magnitude, given by $\delta/\sqrt{\Reyn}\ll \delta$, $\delta$, and $\Reyn^{1/(3 \mu)} \delta \gg \delta$, respectively. Since the thickness of the outer transition layer is much larger than the size of the collision region, a two-level composite expansion could in principle provide the inlet boundary conditions needed for computation of the structure of the stagnation-flow-like collision region. Nevertheless, the three-level expansion can serve the same purpose with higher accuracy and was used in the computations instead.

 
The collision region has been described earlier for vortex flows relevant to tornado phenomena. The early control-volume analysis of \cite{head1977} employed the velocity profiles of \cite{Burggraf.etal.1971} for the lateral incoming-flow boundary condition along with a prescribed form of the outlet velocity profile of the rising core to generate an approximate description. Inviscid solutions were determined by \cite{rotunno1980vorticity} using simple presumed functional forms for the radial and azimuthal velocity distributions across the incoming near-wall boundary layer. Additional inviscid results were obtained by \cite{Fiedler.Rotunno.1986} employing instead tabulated values of the terminal velocity profiles obtained by \cite{Burggraf.etal.1971}. The latter analysis, pertaining to the case $\sigma=0$, focused on computation of the velocity profile approached by the deflected stream above the collision region, which was used to assess the occurrence of vortex breakdown by application of Benjamin's criterion \citep{benjamin1962theory}. Also of interest is the numerical work of \cite{Wilson.Rotunno.1986}, who employed as boundary conditions the velocity profiles measured experimentally in a vortex chamber \citep{1981PhDT160B}. For the moderately large Reynolds number of the experiments, good agreement was found between the inviscid description and the results of numerical integrations of the full Navier-Stokes equations, thereby supporting the idea that the structure of the collision region is fundamentally inviscid. 

\subsection{The rescaled problem}

In the non-slender collision region, of characteristic size $\delta=a/\sqrt{\Reyn}$, all three velocity components have comparable magnitudes $u^* \sim v^* \sim w^* \sim \Gamma/\delta$. Correspondingly, the analysis of this region necessitates introduction of rescaled velocity components $\tilde{u}=u^*/(\Gamma/\delta)$, $\tilde{v}=v^*/(\Gamma/\delta)$, and $\tilde{w}=w^*/(\Gamma/\delta)$ along with a rescaled radial coordinate $\tilde{r}=r^*/\delta$, while the accompanying vertical coordinate $\tilde{z}=z^*/\delta=z$ is that used in the boundary-layer analysis. With these scales, the steady, axisymmetric continuity and momentum equations take the dimensionless form
\begin{align}
\frac{1}{\tilde{r}} \frac{\p }{\p \tilde{r}}(\tilde{r} \tilde{u})+\frac{\p \tilde{w}}{\p \tz}&=0, \label{cont} \\
\tilde{u} \frac{\p \tilde{u}}{\p \tilde{r}}-\frac{\tilde{v}^2}{\tilde{r}}+\tilde{w} \frac{\p \tilde{u}}{\p \tz}&=-\frac{\p \tilde{p}}{\p \tilde{r}}+\frac{1}{\Reyn} \left[ \frac{\p}{\p \tilde{r}} \left( \frac{1}{\tilde{r}} \frac{\p}{\p \tilde{r}}(\tilde{r} \tilde{u}) \right)+\frac{\p^2 \tilde{u}}{\p \tz^2}\right], \label{momr}\\
%
\tilde{u} \frac{\p}{\p \tilde{r}} (\tilde{r} \tilde{v}) +\tilde{w} \frac{\p }{\p \tz}(\tilde{r}\tilde{v})&=\frac{1}{\Reyn} \left[\tilde{r} \frac{\p}{\p \tilde{r}} \left( \frac{1}{\tilde{r}} \frac{\p}{\p \tilde{r}}(\tilde{r} \tilde{v}) \right)+\frac{\p^2 }{\p \tz^2}(\tilde{r} \tilde{v})\right], \label{momtheta} \\
\tilde{u} \frac{\p \tilde{w}}{\p \tilde{r}}+\tilde{w} \frac{\p \tilde{w}}{\p \tz}&=-\frac{\p \tilde{p}}{\p \tz}+\frac{1}{\Reyn} \left[\frac{1}{\tilde{r}}\frac{\p}{\p \tilde{r}} \left( \tilde{r}\frac{\p \tilde{w}}{\p \tilde{r}} \right)+\frac{\p^2 \tilde{w}}{\p \tz^2}\right], \label{momz}
\end{align}
where $\tilde{p}$ denotes the spatial pressure variations scaled with the characteristic dynamic pressure $\rho (\Gamma/\delta)^2$. The distribution of radial and azimuthal velocity at large radial distances is given by the terminal profiles~\eqref{thecomposite} written in terms of the rescaled variables. The solution depends on the Reynolds number $\Reyn =\Gamma/\nu$, which appears explicitly in the equations, and on the ambient swirl level, through the parameter $\sigma$ present in the boundary velocity profiles~\eqref{thecomposite}.

For the large values $\Reyn \gg 1$ considered here, the flow can be expected to be nearly inviscid, although rotational, with viscous effects largely confined to a near-wall layer and to a near-axis core region, both of characteristic size $\delta_v=\delta/\sqrt{\Reyn} \ll \delta$. The inviscid solution is to be investigated in detail below for different values of $\sigma$. To check for consistency of the large-Reynolds-number structure, additional attention is given to the accompanying near-wall boundary layer. The computations reveal that boundary-layer separation occurs at a finite distance $r^* \sim \delta$ regardless of the value of $\sigma$, indicating that the description of the corner region should, in principle, account for viscous effects. This finding motivates additional Navier-Stokes computations for moderately large values of $\Reyn$, which allow us to investigate the extent of the separation region and its dependence on the Reynolds number.

The results to be presented below, extending the previous studies by using boundary velocity profiles that are directly relevant to fire-whirl applications, pertain to cold flow only. Before proceeding with the analysis, it is worth discussing the relevance of the results in connection with localized fire whirls. If one considers, for definiteness, the case of fire whirls developing above liquid-fuel pools, then the fuel-pool diameter $D$ emerges as relevant characteristic length, to be compared with the size of the collision region $\delta$. In the relevant distinguished limit $D \sim \delta$ the fire whirl develops in the collision region, driven by the approaching boundary-layer profile described in~\eqref{thecomposite}. Since $\Reyn \gg 1$, the flame, developing from the liquid-pool rim, would be confined initially to the viscous layer $z^* \sim \delta_v=\delta/\sqrt{\Reyn}$ found in the immediate vicinity of the pool surface and, upon flow deflection near the origin, to the near-axis viscous core found at $r^* \sim \delta/\sqrt{\Reyn}$. The inviscid results given below provide in this case the velocity profile found outside the thin reactive regions as well as the associated imposed pressure gradient, both along the liquid-pool surface and along the vertical axis. The fire whirl, driven by the fast upward flow resulting from the deflection, would continue to develop vertically over distances larger than $\delta$, eventually transitioning to a turbulent plume. Depending on the flow conditions, buoyancy effects, which eventually drive the turbulent plume, can become important already in the reactive boundary layer developing near the liquid-pool surface, possibly helping to prevent boundary-layer separation. A recent attempt to describe this layer \citep{Li.etal.2019} has employed a constant-density model along with the self-similar velocity profile computed from~\eqref{psi_0eq}. Clearly, more accurate numerical computations, accounting for variable-density and buoyancy effects and using as boundary condition the wall-velocity distribution obtained in the inviscid analysis of the collision region, are worth pursuing in future work. 

\subsection{The reduced inviscid formulation}

As first shown by \citet{Hicks.1899}, the inviscid equations that follow from removing the viscous terms involving the factor $\Reyn^{-1}$ from the momentum equations~\eqref{cont}--\eqref{momz} can be combined into a single equation for the stream function $\tp$. As explained by~\citet{batchelor2000introduction}, the development uses the condition that the circulation per unit azimuthal angle $\tilde{C}=\tilde{r} \tilde{v}$ and the total head $\tilde{H}=\tilde{p}+(\tilde{u}^2+\tilde{v}^2+\tilde{w}^2)/2$ remain constant along any given streamline, allowing the azimuthal component of the vorticity to be written in the form
\beq
\frac{\p \tu}{\p \tz}-\frac{\p \tw}{\p \tilde{r}}=\frac{\tilde{C}}{\tilde{r}} \frac{{\rm d} \tilde{C}}{{\rm d} \tp}-\tilde{r} \frac{{\rm d} \tilde{H}}{{\rm d} \tp},
\eeq
finally yielding
\begin{equation} \label{Hicks_eq}
     \pfr{^2\tp}{\tr^2} - \frac{1}{\tr}\pfr{\tp}{\tr} + \pfr{^2\tp}{\tz^2} = - \tilde{C}\frac{\dd\tilde{C}}{\dd\tp}+\tr^2 \frac{\dd \tilde{H}}{\dd\tp},
\end{equation}
upon substitution of the expressions $\tilde{r} \tilde{w}=\p \tp /\p \tilde{r}$ and $\tilde{r} \tilde{u}=-\p \tp /\p \tz$. As shown below, the functions $\tilde{C}(\tp)$ and $\tilde{H}(\tp)$ are to be evaluated using the terminal velocity profiles~\eqref{thecomposite} written in the simplified form $\tr \tu=F_0(\tz)$ and $\tr \tv=G_0(\tz)$, corresponding to $\Reyn \rightarrow \infty$, with the functions $F_0$ and $G_0$ carrying the dependence on $\sigma$, as shown in figure~\ref{fig:G0F0}.

Since the streamlines lie parallel to the wall as $\tr \to \infty$, the equation $\tilde{r} \tilde{u}=-\p \tp /\p \tz$ provides $\tp=-\int_0^{\tz} F_0 {\rm d} \tz$, which can be used to determine the boundary distribution
\beq \label{phi_infty}
\tp=\tp_\infty(\tz)=-\int_0^{\tz} F_0 {\rm d} \tz \quad {\rm as} \; \tr \to \infty,
\eeq
and, implicitly through
\beq \label{z_infty}
\tp=-\int_0^{\tz_\infty} F_0 {\rm d} \tz,
\eeq
the height $\tz_\infty(\tp)$ at which a given streamline originates. While the head tends to a uniform value as the velocity decays far from the axis, so that ${\rm d} \tilde{H}/{\rm d} \tp=0$ in~\eqref{Hicks_eq}, the circulation $\tilde{C}$ varies between streamlines, yielding a contribution to~\eqref{Hicks_eq} that can be evaluated by using $\tilde{C}\dd\tilde{C}/\dd\tp=F'_0[\tz_\infty(\tp)]$, derived with use of~\eqref{eulersol}. The problem then reduces to that of integrating the nonlinear equation
\begin{equation} \label{Hicks_eq2}
     \pfr{^2\tp}{\tr^2} - \frac{1}{\tr}\pfr{\tp}{\tr} + \pfr{^2\tp}{\tz^2} = m(\tp),
\end{equation}
where $m(\tp)=-F'_0[\tz_\infty(\tp)]$,
with boundary conditions
\begin{align} \label{Hicks_bc2}
    \tp(0,\tz) = \tp(\tr,0)=\tp(\infty,\tz) -\tp_\infty(\tz) =0 \quad \text{and} \quad  \pfr{\tp}{\tz} = 0 \quad \text{as} \quad \tz\rightarrow \infty.
\end{align}
The solution depends on $\sigma$ through the derivative and antiderivative of the function $F_0$, which appear on the right-hand side of~\eqref{Hicks_eq2} and in the boundary distribution $\tp_\infty$ given in~\eqref{phi_infty}, respectively. Note that, despite the slow algebraic decay $F_0 \simeq -C_{2} \tz^{-\mu}$ indicated in~\eqref{largez}, the condition $\mu > 1$ guarantees that the antiderivative $\int_0^{\tz} F_0 {\rm d} \tz$ approaches a finite value as $\tz \rightarrow \infty$ for all values of $\sigma$. Once $\tp(\tr,\tz)$ has been determined, the distribution of azimuthal velocity can be evaluated with use of $\tr \tv=\tilde{C}(\tp)=G_0[\tz_\infty(\tp)]$ supplemented by~\eqref{z_infty}, as follows in the inviscid limit from conservation of circulation along streamlines. This reduced description is the basis of many of the early vortex-core studies \citep{rotunno1980vorticity,Fiedler.Rotunno.1986,Wilson.Rotunno.1986}.

\begin{figure}
\centering
\includegraphics[width=0.9\textwidth]{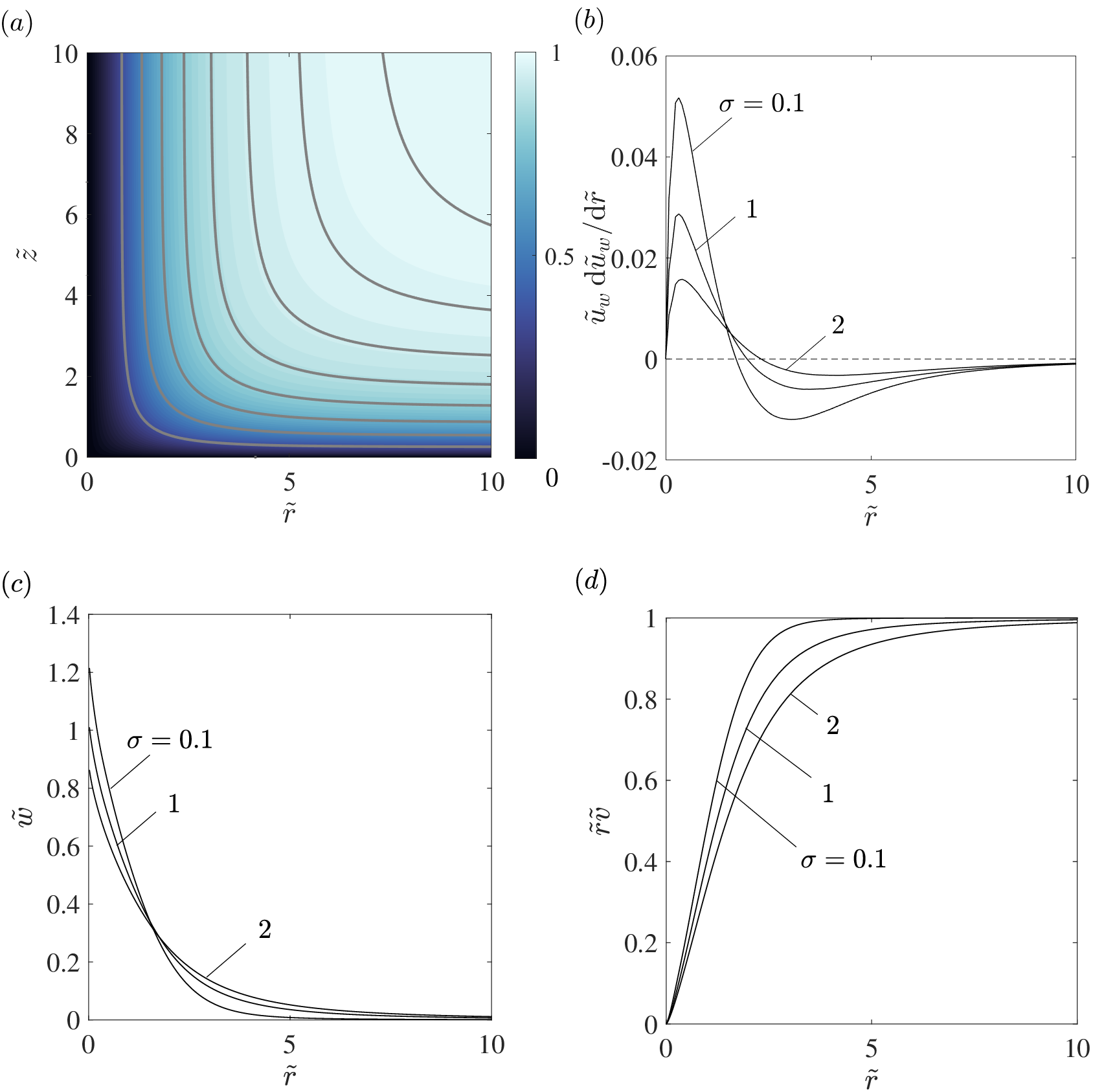}
\caption{The inviscid structure of the collision region calculated from \eqref{Hicks_eq2}, including the streamlines $\tp=(0.25,0.5,0.75,1.0,1.25,1.5,1.75,2.0)$ (solid curves) and circulation $\tilde{r} \tilde{v}$  (color map) for $\sigma = 1$ (a), the negative radial pressure gradient on the wall obtained from $\tu_w(\tr) =- (1/\tr)\pfi{\tp}{\tz}|_{\tz=0}$ for $\sigma=(0.1,1.0,2.0)$ (b), and the corresponding profiles of axial velocity $\tilde{w}$ (c) and circulation $\tilde{r} \tilde{v}$ (d) approached as $\tilde{z}\rightarrow \infty$.}
    \label{fig:iviscidcollision}
\end{figure}

\subsection{Sample results}

Numerical solutions to the problem defined in~\eqref{Hicks_eq2} and~\eqref{Hicks_bc2} were
obtained using a finite-element method \citep{Hecht.2012}. No convergence problems were
encountered for any value of $\sigma$. The streamlines and circulation distribution
corresponding to $\sigma = 1$ are shown in figure~\ref{fig:iviscidcollision}(a). The integration
provides in particular the slip velocity along the wall $\tu_w = -
(1/\tr)\pfi{\tp}{\tz}|_{\tz=0}$ and the associated radial pressure gradient $-\p \tilde{p}/\p
\tilde{r}=\tu_w {\rm d}\tu_w/ {\rm d} \tilde{r}$, with the latter shown in
figure~\ref{fig:iviscidcollision}(b) for selected values of $\sigma$. In all cases, the pressure
gradient, whose magnitude increases with decreasing $\sigma$, is favorable far from the axis and
adverse near the axis. The value of $\tu_w {\rm d}\tu_w/ {\rm d} \tilde{r}$ is seen to decrease
linearly with the radial distance on approaching the origin, a behavior that is consistent with
the local stagnation-point solution $\tp \sim\tr^2\tz$ that prevails at $\tr^2+\tz^2\ll 1$, as
follows from a local analysis of \eqref{Hicks_eq2}. 


The deflected streamlines become aligned with the axis for $\tilde z \gg 1$, when the stream
function approaches the limiting distribution $\Psi(\tr)=\tp(\tr,\infty)$, to be determined from
integration of
\begin{equation}  \label{Psi_problem}
\Psi''-\Psi'/\tr=m(\Psi), \quad \Psi(0)=\Psi(\infty)+\int_0^\infty F_0 {\rm d}\tilde z=0,
\end{equation} 
the one-dimensional counterpart of~\eqref{Hicks_eq2}. The corresponding distributions of axial
velocity $\tilde{w}=\Psi'/\tr$ and circulation $\tr \tilde{v}=\tilde C[\Psi(\tr)]$, which
provide the initial conditions for studying the development of the flow above the collision
region, are shown in figures~\ref{fig:iviscidcollision}(c) and~\ref{fig:iviscidcollision}(d) for
the three representative flow inclinations $\sigma =(0.1,1,2)$, for which the boundary values of
the stream function are $\Psi(\infty)=(1.62,4.16,5.57)$, respectively. As can be seen, the
rising jet is found to be wider for increasing $\sigma$, a consequence of the shape of the
boundary velocity distributions $F_0$ and $G_0$. The integration provides, in particular, the
peak axial velocity $\tilde{w}_0=\tilde{w}(0)$, given by $\tw_0=(1.22,1.01,0.86)$ for $\sigma
=(0.1,1,2)$. Near the axis, where $m=-C_1^2 \lambda \Psi^{2\lambda -1}$ and $\tilde{C}=C_1
\Psi^\lambda$ with $\lambda=0.6797$, as follows from~\eqref{smallz}, the solution takes the form 
\begin{equation}
\tilde{w}=\tilde{w}_0-\frac{C_1^2 \lambda \tilde{w}_0^{2 \lambda-1} \tr^{4\lambda-2}}{2^{2\lambda}(2 \lambda-1)} \quad {\rm and} \quad \tr \tilde{v}=C_1 \left(\tilde{w}_0 \tr^2/2 \right)^\lambda.
\end{equation}
Since $\lambda < 3/4$, the axial velocity of the inviscid solution displays an infinite slope at
the axis. This characteristic of the velocity distribution, which would disappear in the
presence of viscous forces, is not revealed in the early results of \cite{Fiedler.Rotunno.1986}
because the tabulated representations of $F_0$ and $G_0$ employed in their description, taken
from \cite{Burggraf.etal.1971}, did not contain enough points to reproduce the near-wall
behavior~\eqref{smallz}.


\subsection{The boundary layer in the collision region}

\begin{figure}
\centering
\includegraphics[width=\textwidth]{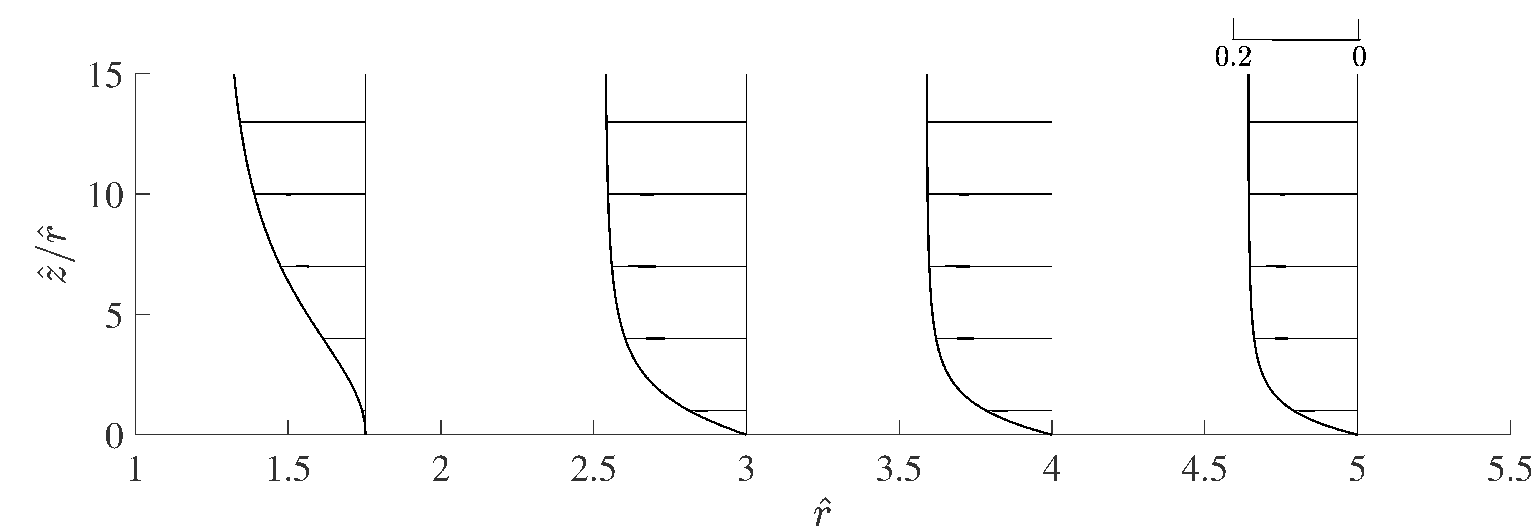}
\caption{Boundary-layer profiles of radial velocity at selected locations obtained by
    integration of~\eqref{bleq_collision} for $\sigma = 1$.}
\label{fig:sepblprof}
\end{figure}

The inviscid flow described above is accompanied by a near-wall viscous boundary layer with
characteristic thickness $z^*\sim \delta_v=\delta/\sqrt{\Reyn}$ at radial distances $r^* \sim
\delta$. As seen in figure~\ref{fig:iviscidcollision}(b) this boundary layer develops under the
action of a pressure gradient $-\p \tilde{p}/\p \tilde{r}=\tu_w {\rm d}\tu_w/ {\rm d} \tilde{r}$
that is negative (favorable) at large radial distances but becomes positive (adverse) on
approaching the axis. Clearly, the validity of the inviscid solution as a representation of the
flow for $\Reyn \gg 1$ requires that the boundary layer remains attached, that being the
assumption underlying previous
descriptions~\citep{Fiedler.Rotunno.1986,Wilson.Rotunno.1986,Rotunno.2013}. Examination of this
aspect of the problem requires introduction of the rescaled variables
\begin{equation}
    \hr = \frac{r^*}{\delta}= \tilde r, \quad \hz = \frac{z^*}{\delta_v}=\sqrt{\Reyn}\tilde z, \quad \hu = \frac{u^*}{\Gamma/\delta}= \tu,  \quad \hw = \frac{\sqrt{\Reyn}w^*}{\Gamma/\delta}= \sqrt{\Reyn} \tilde w
\end{equation}
to write the boundary-layer equations
\begin{align}
    \hat u\pfr{\hat u}{\hat r} +\hat w\pfr{\hat u}{\hat z}  = \tu_w \frac{\dd \tu_w}{\dd \hat r}  + \pfr{^2 \hat u}{ \hat z^2}, \label{bleq_collision}\\
    \pfr{}{\hat r}(\hat r \hat u) + \pfr{}{\hat z}(\hat r \hat w) =0,
\end{align}
and associated initial and boundary conditions
\beq \label{blbc_collision}
\begin{split}
&\hr \to \infty:\quad \hr \hu = -\psi'_0(\hat z/\hat r), \\ 
&\hz = 0: \quad \hu = \hw = 0, \quad \hz \to \infty:\quad \hu \to \tu_w(\hr),
\end{split}
\eeq
involving the apparent slip velocity $\tu_w$ of the inviscid collision region, which carries the
dependence on $\sigma$, and on the rescaled stream function $\psi_0$ across the viscous
sublayer, determined from~\eqref{psi_0eq}.

Numerical integration of~\eqref{bleq_collision}--\eqref{blbc_collision} for decreasing values of
$\hr$ reveals that for all $\sigma$ the boundary layer separates at a radial location $\hr \sim
1$, where the velocity profile develops an inflection point at the wall, preventing integration
beyond that point. Illustrative results are shown in figure~\ref{fig:sepblprof} for
$\sigma=1$, for which separation is predicted to occur at $\hr \simeq 1.76$.

\subsection{The viscous structure of the collision region}

\begin{figure}
\centering
\includegraphics[width = \textwidth]{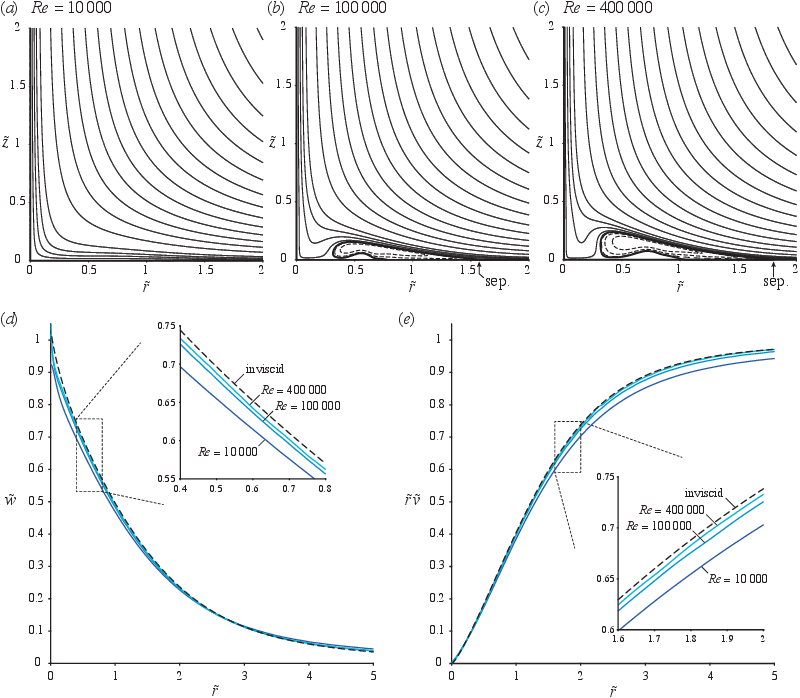}
\caption{The viscous structure of the collision region for $\sigma=1$, including
    streamlines obtained by integration of~\eqref{cont}--\eqref{momz} for (a) $\Reyn = 10^4$,
    (b) $\Reyn = 10^5$, and (c) $\Reyn = 4\times10^5$, together with a comparison of radial
    profiles of (d) axial velocity $\tilde{w}$ and (e) circulation
    $\tilde{r}\tilde{v}$ at $\tilde{z} = 5$ with the inviscid results shown in
    figure~\ref{fig:iviscidcollision}(c) and \ref{fig:iviscidcollision}(d).
    Streamlines in (a)--(c) are not equispaced. The arrows in (b) and (c) indicate the location
    where the boundary layer separates.}
\label{fig:separation}
\end{figure}


The predicted separation of the boundary layer, questioning the validity of the inviscid description, was further investigated numerically by integrating
the complete Navier-Stokes equations~\eqref{cont}--\eqref{momz} with a
Newton-Raphson method in combination with the finite-element solver FreeFem++ \citep{Hecht.2012}
for increasing values of $\Reyn$ and different values of $\sigma$. A cylindrical computational
domain with outer radius $\tr_{\rm max} \gg 1$ and height $\tz_{\rm max} \gg 1$ was employed in
the integrations. The three-level composite expansions of \eqref{thecomposite}, written in terms
of the collision-region variables, were used to provide the inlet boundary conditions at
$\tr=\tr_{\rm max}$. Additional boundary conditions include $\tu=\tv=\tw=0$ at $\tz=0$,
$\tu=\tv=\p \tw/\p \tr=0$ at $\tr=0$, and the outflow condition $\partial p/\partial z = 0$
at $\tz=\tz_{\rm max}$. The
results for the flow in the collision region $\tr \sim \tz \sim 1$ were found to be independent
of the size of the computational domain provided that the boundaries were selected in the ranges
$5 \ltsim \tr_{\rm max} \ltsim 10$ and $10 \ltsim \tz_{\rm max} \ltsim 20$. Illustrative results
using $\tr_{\rm max}=8$ and $\tz_{\rm max}=16$ are shown in figure~\ref{fig:separation}(a), (b)
and (c) for $\sigma=1$ and three different values of the relevant Reynolds number $\Reyn$. It should be noted that previous experimental results \citep{phillips1985vortex} suggest that, for the two largest Reynolds numbers considered, namely, $\Reyn = 10^5$ and $\Reyn = 4\times10^5$, the boundary layer of the steady-flow solutions considered here is probably unstable and would experience transition to a turbulent state, but that aspect of the problem is not investigated in our numerical computations, which are focused instead on the emergence of boundary-layer separation.

It can be seen in figure~\ref{fig:separation}(a) that the structure of the flow for $\Reyn=10^4$ is very similar to
the inviscid structure, in that the boundary layer remains attached and the resulting
streamlines are similar to those shown in
figure~\eqref{fig:iviscidcollision}(a). By way of contrast, the flow structure found when the
Reynolds number is increased to $\Reyn=10^5$ is markedly different
(figure~\ref{fig:separation}(b)). The streamline pattern
reveals the presence of a slender recirculating bubble adjacent to the wall, generated by the
separation of the boundary layer at $\tr \simeq 1.55$, and subsequent reattachment at $\tr \simeq
0.3$. Further increasing the Reynolds number to $\Reyn = 4\times10^5$ causes the recirculation
bubble to enlarge, and this also moves the location of the point at which the boundary layer separates to
$\tilde{r} = 1.75$, approaching the value $\tilde{r} = 1.76$ predicted by the boundary-layer computations.

The closed recirculating bubble has a limited effect on the vertical jet issuing from the collision region. This is quantified in
figures~\ref{fig:separation}(d) and (e), where profiles of axial velocity and circulation
at $\tilde{z} = 5$ for the three values of the Reynolds number considered before are compared
with the inviscid results shown in figure~\ref{fig:iviscidcollision}(c) and (d). A noticeable difference is found near the axis, where the sharp peak of the inviscid axial velocity is smoothed by the viscous forces. The resulting near-axis boundary layer is thicker for smaller Reynolds numbers, resulting in a smaller peak velocity. The quantitative agreement everywhere else is quite satisfactory, with the viscous results approaching the inviscid profile for increasing values of the Reynolds number.

It is worth pointing out that closed recirculating bubbles, similar to those shown in figures~\ref{fig:separation}(b) and~\ref{fig:separation}(c), were observed near the end wall in early flow simulations of Ward-type vortex chambers with Reynolds number $10^3$ (based on the flow rate and on the vortex-chamber radius) when the swirl level was sufficiently low \citep{rotunno1979study}. The separation bubble disappears for increasing swirl level  \citep{rotunno1979study} and is not present in subsequent computations of the same flow at Reynolds number $10^4$ \citep{Wilson.Rotunno.1986}, for which the flow was shown to be fundamentally inviscid. No indication of boundary-layer separation was found in recent tornado simulations at much higher Reynolds numbers \citep{rotunno2016axisymmetric} employing a prescribed forcing term in the vertical momentum equation to generate the motion. The differences between the results of these previous simulations \citep{rotunno1979study,Wilson.Rotunno.1986,rotunno2016axisymmetric} and the predictions reported above are attributable to the differences in the associated flow field, suggesting that the detailed distribution of near-wall velocity plays a critical role in the occurrence of boundary-layer separation on approaching the axis.

\section{Conclusions}
\label{sec:conclusions}

This investigation clarifies a number of aspects of the structure of the boundary layer that
will develop between an axisymmetric fire and swirl-producing obstacles located at a large but
finite radius from its center, by analyzing situations in which the external inviscid flow can
be described as a superposition of a potential vortex and the Taylor potential flow generated by
a turbulent plume. A one-parameter family of solutions was developed, that parameter being the
ratio of the inward radial component of velocity to the azimuthal (swirl) component at the
cylindrical swirl-generation boundary, thereby extending an earlier, tornado-motivated analysis
(for which that parameter vanishes) to conditions of interest for fire whirls. The initially
non-similar boundary layer evolves, at radii small compared with the radius of the obstacle
location, into a three-level structure composed of an inner self-similar viscous sublayer, below
a thicker, self-similar, rotational, inviscid layer which, in turn, lies below an even thicker,
self-similar, still rotational, inviscid layer of transition to the external potential flow. A
composite expansion is given that describes the structure of this three-level boundary layer,
which helps in addressing computationally the flow near the axis of symmetry, needing study for accurate and
complete descriptions of fire-whirl structures, including their stability and the onset of vortex breakdown. For instance, the composite expansion has been used recently \citep{Carpio.etal.2020} as boundary condition for the numerical description of the structure of fire whirls lifted over liquid-fuel pools, stabilized by vortex breakdown when the level of ambient swirl becomes sufficiently large \citep{Xiao.etal.2016}. Similar numerical investigations can be useful in addressing unsteady fire-whirl dynamics, including transitions between attached and lifted flames and intermittent vortex breakdown, which have been observed in controlled laboratory experiments~\citep{coenen2019observed}.

With decreasing radius, the thickness of the viscous sublayer decreases, and the azimuthal
velocity decreases, while the inward radial velocity increases, leading to a collision region
near the axis, of a size proportional to the square root of the ratio of the kinematic viscosity
to the circulation (the reciprocal of a Reynolds number), in which the flow experiences
transition from predominantly radially inward to predominantly upward motion. This collision
region is described, in general, by the full Navier-Stokes equations, but it develops a
dominantly inviscid structure for large enough Reynolds numbers, with boundary layers at the
base and on the axis. The colliding inward motion produces a stagnation-flow type of behavior,
which results in an unfavorable pressure gradient acting on the viscous base flow, leading to
its separation at high enough Reynolds numbers, but which apparently turns out to be followed by
re-attachment, at least at Reynolds numbers accessible computationally, so that the upward
outflow can be estimated reasonably. These rather complex constant-density boundary-layer
structures in fire whirls underlie the combustion effects which, by decreasing the gas density,
give rise to the tall fire whirls that generally are seen. Proper complete analyses of these
fire whirls and of the vortex-breakdown phenomena that occur in them at sufficiently small
values of the ratio of radial to azimuthal incoming velocity need to take into account the flow
characteristics uncovered in the present work. 

\begin{acknowledgments}
This work was supported by the National Science Foundation through grant \#	1916979.
\end{acknowledgments}

\bibliographystyle{jfm}
\bibliography{references}

\end{document}